\documentstyle[12pt,epsfig]{article}
\renewcommand{\section}[1]{\vspace{6pt} \noindent\mbox{#1} \newline \noindent}
\renewcommand{\subsection}[1]{\vspace{6pt} \noindent\mbox{\underline{#1}} 
\newline \noindent}
\renewcommand{\subsubsection}[1]{\vspace{6pt} \noindent\mbox{\underline{#1}}
\noindent}

\newfont{\sansb}{cmssbx10}
\newfont{\sans}{cmss10}

\begin{document}
{\small OG 4.1.19 \vspace{-24pt}\\}
{\center \bf \Large VERY HIGH ENERGY GAMMA RAYS FROM THE VELA PULSAR/NEBULA
\vspace{6pt}\\}
T.~Yoshikoshi$^1$, T.~Kifune$^1$, S.A.~Dazeley$^2$, P.G.~Edwards$^{2,3}$,
T.~Hara$^4$, Y.~Hayami$^5$, S.~Kamei$^5$, R.~Kita$^6$, T.~Konishi$^7$,
A.~Masaike$^8$, Y.~Matsubara$^9$, Y.~Mizumoto$^{10}$, M.~Mori$^{11}$,
H.~Muraishi$^6$, Y.~Muraki$^9$, T.~Naito$^{12}$, K.~Nishijima$^{13}$,
S.~Ogio$^5$, J.R.~Patterson$^2$, M.D.~Roberts$^1$, G.P.~Rowell$^2$,
T.~Sako$^9$, K.~Sakurazawa$^5$, R.~Susukita$^{14}$, A.~Suzuki$^7$,
R.~Suzuki$^5$, T.~Tamura$^{15}$, T.~Tanimori$^5$, G.J.~Thornton$^{1,2}$,
S.~Yanagita$^6$ and T.~Yoshida$^6$ \vspace{6pt}\\
{\it
$^1$Institute for Cosmic Ray Research, University of Tokyo, Tokyo 188, Japan\\
$^2$Department of Physics and Mathematical Physics, University of Adelaide,
  South Australia 5005, Australia\\
$^3$Institute of Space and Astronautical Science, Kanagawa 229, Japan\\
$^4$Faculty of Management Information, Yamanashi Gakuin University,
  Yamanashi 400, Japan\\
$^5$Department of Physics, Tokyo Institute of Technology, Tokyo 152, Japan\\
$^6$Faculty of Science, Ibaraki University, Ibaraki 310, Japan\\
$^7$Department of Physics, Kobe University, Hyogo 637, Japan\\
$^8$Department of Physics, Kyoto University, Kyoto 606, Japan\\
$^9$Solar-Terrestrial Environment Laboratory, Nagoya University, Aichi 464,
  Japan\\
$^{10}$National Astronomical Observatory of Japan, Tokyo 181, Japan\\
$^{11}$Department of Physics, Miyagi University of Education, Miyagi 980,
  Japan\\
$^{12}$Department of Earth and Planetary Physics, University of Tokyo,
  Tokyo 113, Japan\\
$^{13}$Department of Physics, Tokai University, Kanagawa 259, Japan\\
$^{14}$Institute of Physical and Chemical Research, Saitama 351-01, Japan\\
$^{15}$Faculty of Engineering, Kanagawa University, Kanagawa 221, Japan
  \vspace{-12pt}\\}
{\center ABSTRACT\\}
We have observed the Vela pulsar region at TeV energies using the 3.8~m
imaging \v{C}erenkov telescope near Woomera, South Australia every year
since 1992. This is the first concerted search for pulsed and unpulsed
emission from the Vela region, and the
imaging technique also allows the location of the emission within
the field of view to be examined. A significant excess of gamma-ray-like
events is found offset from the Vela pulsar to the southeast by
about $0^\circ.13$. The excess shows the behavior expected of gamma-ray
images when the asymmetry cut is applied to the data. There is no evidence
for the emission being modulated with the pulsar period -- in contrast
to earlier claims of signals from the Vela pulsar direction.

\setlength{\parindent}{1cm}
\section{INTRODUCTION}
The Vela pulsar, PSR~B$0833-45$, is an 89~msec pulsar and the brightest
100~MeV gamma-ray pulsar. The Crab pulsar and PSR~B$1706-44$ are also
strong 100~MeV gamma-ray sources, and both have been detected as
very high energy (VHE) gamma-ray sources (e.g. Weekes et al., 1989;
Tanimori et al., 1994; Kifune et al., 1995). The emission from
these VHE sources is apparently unpulsed and the acceleration site
for the progenitor electrons is probably the shock generated
by the pulsar wind colliding with circumstellar matter. There have been some
claims\footnote{Their results were not confirmed by later observations
(see Edwards et al., 1994).} for pulsed VHE emission from the Vela pulsar
(Grindlay et al., 1975; Bhat et al., 1987). However, with the exception
of a short subset of Grindlay et al.'s data, only pulsed emission
has been searched for in the previous VHE observations,
and unpulsed gamma rays are likely emitted also from the Vela pulsar/nebula
owing to a mechanism presumably similar as the Crab pulsar/nebula.
The atmospheric \v{C}erenkov imaging technique has made possible
the search for unpulsed VHE gamma rays, and the CANGAROO 3.8~m telescope
(Hara et al., 1993) is used for the first attempt of searching
for unpulsed VHE emission from the Vela pulsar/nebula.

We have observed the Vela pulsar region by the CANGAROO 3.8~m telescope
every year since 1992 and usable data were obtained in 1993, 1994, 1995
and 1997. The on-source data from 1993 to 1995 amount to about 174~hours
in total, of which about 119~hours data remain after rejecting the data
affected by clouds. Almost the same amount of off-source data has been taken
night by night. The 3.8~m mirror was recoated in October 1996
and the reflectivity was improved from $\sim 45$~\% to $\sim 90$~\% at 480~nm
(Wild et al., 1995). The 1997 data are still under analysis.
Only the 1993 to 1995 data are used in the following analyses.

\section{ANALYSES AND RESULTS}
The analysis method is based on parameterization of the shape, location
and orientation of \v{C}erenkov images detected by an imaging camera
(Hillas, 1985; Reynolds et al., 1993). The gamma-ray selection criteria
in the present analysis are determined by Monte Carlo simulations to be
$0^\circ.01 < {\rm width} < 0^\circ.09$,
$0^\circ.1 < {\rm length} < 0^\circ.4$, $0.3 < {\rm conc}$,
$0^\circ.7 < {\rm distance} < 1^\circ.2$ and ${\rm alpha} < 10^\circ$.
The simulations show that more than 99~\% of background events are rejected
after the above selections are made, while $\sim 50$~\% of gamma-ray images
from a point source remain.

First, we searched for a gamma-ray signal from the pulsar position
and found a significant excess of on-source events above the background
(off-source) level. However, the peak profile of the excess distribution
against alpha was broader than expected from a point source.
Next, gamma-ray emission was scanned from other positions than
the pulsar position. The position of the maximum excess counts was
found to be offset from the pulsar by about $0^\circ.13$.
Figure~1 shows the alpha distributions of the events from the position
of the maximum excess counts after the other selection criteria have
been applied.
\begin{figure}
\epsfig{file=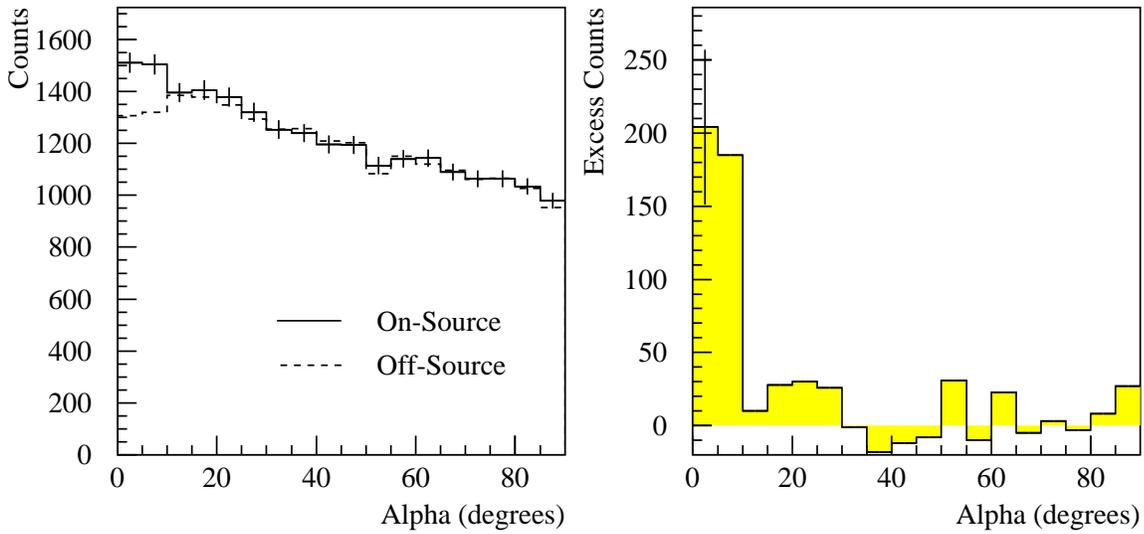,width=17cm}
\caption{\it Alpha distributions of gamma-ray-like events in
the direction of the maximum excess counts, which is offset by about
$0^\circ.13$ to the southeast from the Vela pulsar, and corresponds to
the peak position in Figure~2. The on-source and off-source distributions
are shown in the left. The excess counts of the on-source
above the background (off-source) level are plotted in the right.}
\end{figure} 
The statistical significance of the excess after including the selection
of alpha is at the 5.2~$\sigma$ level.

In the usual image analysis, the image parameters are calculated
assuming gamma-ray images to have a simple elliptical shape.
Gamma-ray images, however, are asymmetrical along their major axes,
having an elongated, comet-like shape with the ``tail'' pointing away
from the source position. This feature can be characterized and quantified
by another image parameter called the ``asymmetry'' (Punch, 1993).
From Monte Carlo simulations, about 80~\% of gamma-ray images have
positive values of ``asymmetry'', while the distribution of background images
is almost symmetrical because of their isotropic arrival directions
in the field of view (Yoshikoshi et al., 1997). The asymmetry parameter
was calculated assuming that the source lies at the position
of the maximum excess counts, to examine the gamma-ray-like feature
of the excess events. The asymmetry distribution of the excess events
was found to be asymmetrical with the peak on the positive side
as expected from the simulation. This result gives clear corroborative
evidence that the detected excess events of Figure~1 are truly
due to gamma rays. The significance of 5.2~$\sigma$ for the excess counts
in the alpha distribution of Figure~1 increases to 5.8~$\sigma$
by adding ${\rm asymmetry} > 0$ to the gamma-ray selection criteria.

As an alternative method to calculating the alpha parameter, it is possible
to infer and assign the true direction of a gamma ray from the location and
orientation of its observed \v{C}erenkov image. We have calculated
a probability density for the true direction for each observed
gamma-ray image using Monte Carlo simulations (Yoshikoshi, 1996).
Thus, a density map of gamma-ray directions can be plotted against
right ascension and declination by adding up all of the probability densities
of the gamma-ray-like images. The contribution of the gamma-ray source
in the field of view can then be found as an event excess in the on-source map
over the background (off-source) map. Figure~2 shows the density map
of the excess events for the Vela pulsar data plotted as a function
of right ascension and declination.
\begin{figure}
\epsfig{file=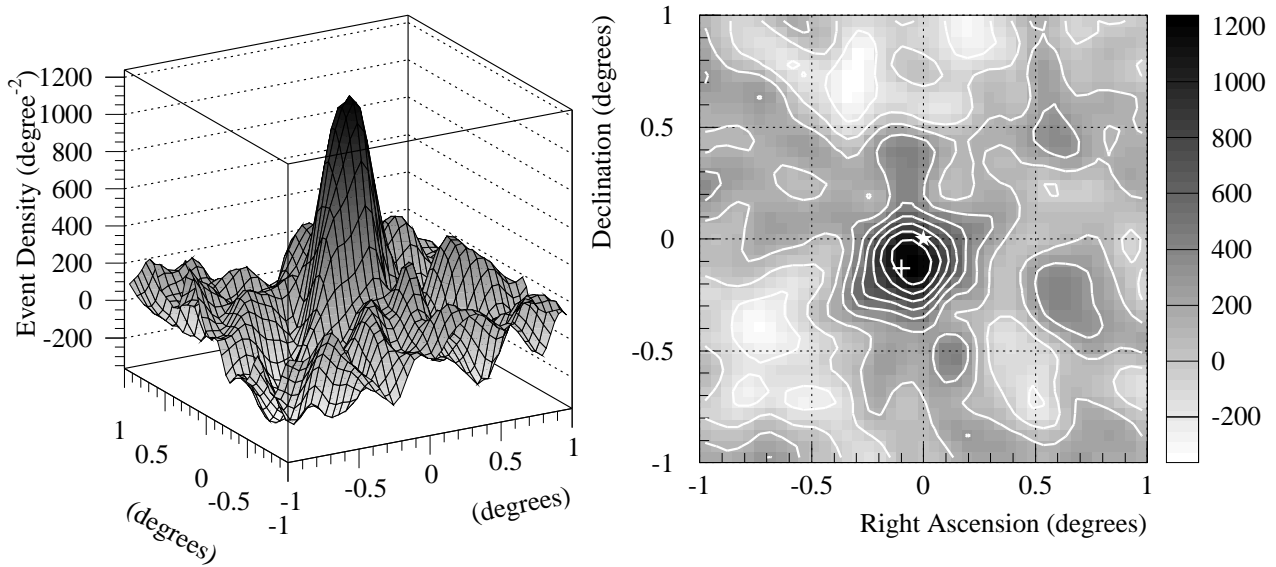,width=17cm}
\caption{\it A density map of excess counts around the Vela pulsar
plotted as a function of right ascension and declination. North is up and
east is to the left in the contour map, and the gray scale at the right
is in $\it counts\; degree^{-2}$. The ``star'' at the origin of
the contour map indicates the position of the Vela pulsar. The position of
the maximum excess counts from the 1997 data is indicated by the ``cross''.}
\end{figure} 
The significant excess due to a gamma-ray source near the pulsar exists
in this map. The position of maximum emission is offset from the pulsar,
which is indicated by the ``star'' mark in the contour map, to a position
southeast by about $0^\circ.13$. The alpha distribution of Figure~1
is plotted for the position of maximum emission. 
We have found a gamma-ray-like event excess also from the 1997 data with
a significance of more than 4~$\sigma$ and the position of
maximum excess counts, indicated by the ``cross'' in Figure~2,
agrees with that from the 1993 to 1995 data within the statistical error
($\sim 0^\circ.04$).

The gamma-ray integral flux calculated for the position of
the maximum excess counts is
$(2.9 \pm 0.5 \pm 0.4) \times 10^{-12}\; \rm photons\; cm^{-2}\; sec^{-1}$
above 2.5~TeV, where the first and second errors are statistical and
systematic respectively. The threshold energy, which we defined as the energy
of the maximum differential flux of detected gamma-ray-like events,
has been estimated from Monte Carlo simulations to be 2.5~TeV,
assuming a power law spectrum with a photon index of $-2.5$.
The systematic error for this threshold energy is about $\pm 1.0$~TeV,
which is due to uncertainties in the photon index,
the trigger conditions and the reflectivity of the mirror.
The data have been divided to calculate year-by-year fluxes and the fluxes
are consistent with no variation having been detected on the time scale
of two years.

The periodicity of the events detected from the Vela pulsar direction and
the direction of the maximum excess counts has been investigated
using our 1994 and 1995 data for which the GPS receiver was available
(Yoshikoshi et al., 1997). No significant value of the $Z_2^2$ statistic
(Buccheri et al., 1983) has been found for either the excess direction
nor the pulsar direction.

\section{CONCLUSION}
We have observed the Vela pulsar region with the CANGAROO 3.8~m telescope
since 1992 and found a VHE gamma-ray signal from the direction $0^\circ.13$
offset from the Vela pulsar to the southeast at the 5.8~$\sigma$ level
in the 1993 to 1995 data. This gamma-ray signal corresponds
to the integral flux of
$(2.9 \pm 0.5 \pm 0.4) \times 10^{-12}\; \rm photons\; cm^{-2}\; sec^{-1}$
above 2.5~TeV and is consistent with steady emission over the two years.
No pulsed emission modulated with the pulsar period has been detected.

\section{ACKNOWLEDGMENTS}
This work is supported by International Scientific Research Program of
a Grant-in-Aid in Scientific Research of the Ministry of Education, Science,
Sports and Culture, Japan, and by the Australian Research Council.
T.~Kifune and T.~Tanimori acknowledge the support of the Sumitomo Foundation.
The receipt of JSPS Research Fellowships (P.G.E., T.N., M.D.R., K.S., G.J.T.
and T.~Yoshikoshi) is also acknowledged.

\section{REFERENCES}
\setlength{\parindent}{-5mm}
\begin{list}{}{\topsep 0pt \partopsep 0pt \itemsep 0pt \leftmargin 5mm
\parsep 0pt \itemindent -5mm}
\vspace{-15pt}
\item Bhat, P.N., Gupta, S.K., Ramanamurthy, P.V., et al., {\it A\&A},
      {\bf 178}, 242 (1987).
\item Buccheri, R., Bennett, K., Bignami, G.F., et al., {\it A\&A},
      {\bf 128}, 245 (1983).
\item Edwards, P.G., Thornton, G.J., Patterson, J.R., et al., {\it A\&A},
      {\bf 291}, 468 (1994).
\item Grindlay, J.E., Helmken, H.F., Hanbury Brown, R., et al. {\it ApJ},
      {\bf 201}, 82 (1975).
\item Hara, T., Kifune, T., Matsubara, Y., et al.,
      {\it Nucl. Inst. Meth. Phys. Res. A}, {\bf 332}, 300 (1993).
\item Hillas, A.M., {\it Proc. 19th ICRC}, {\bf 3}, 445 (1985).
\item Kifune, T., Tanimori, T., Ogio, S., et al. {\it ApJ}, {\bf 438},
      L91 (1995).
\item Punch, M., Ph.D. thesis, National University of Ireland (1993).
\item Reynolds, P.T., Akerlof, C.W., Cawley, M.F., et al., {\it ApJ},
      {\bf 404}, 206 (1993).
\item Tanimori, T., Tsukagoshi, T., Kifune, T., et al., {\it ApJ},
      {\bf 429}, L61 (1994).
\item Weekes, T.C., Cawley, M.F., Fegan, D.J., et al., {\it ApJ},
      {\bf 342}, 379 (1989).
\item Wild, N., Dowden, S., Patterson, J.R. and Thornton, G.J.,
      {\it Proc. 24th ICRC}, {\bf 1}, 966 (1995). 
\item Yoshikoshi, T., Kifune, T., Dazeley, S.A., et al., {\it ApJ}, submitted
      (1997).
\item Yoshikoshi, T., Ph.D. thesis, Tokyo Institute of Technology (1996).
\end{list}

\end{document}